\documentclass[proceedings]{JHEP}

\conference{Non-perturbative Quantum Effects 2000}

\title{Black Hole Radiation On and Off the Brane}

\author{Roberto Emparan\thanks{Based on work in
collaboration with G.T.\ Horowitz and R.C.\ Myers \cite{rEHM3}.}\\
        Departamento de F{\'\i}sica Te\'orica\\
Universidad del Pa{\'\i}s Vasco, Apdo.\ 644, E-48080 Bilbao, Spain\\
        E-mail: \email{wtpemgar@lg.ehu.es}}

\abstract{After a brief review of the description of black holes on
branes, we examine the evaporation of a small black hole on a brane in
a world with large extra dimensions. We show that, contrary to previous
claims, most of the energy is radiated into the modes on the brane. This
raises the possibility of observing Hawking radiation in future high
energy colliders if there are large extra dimensions.}

\begin{document}

\section{Introduction}

It has been realized that space may have extra compact dimensions as
large as a millimeter, in a way that can be consistent with all current
observations \cite{rADD}. To achieve this, the Standard Model fields are
required to be confined on a three-brane, and only gravity propagates
into the extra dimensions. The validity of the Standard Model up to
energies around a TeV requires the thickness of the brane to be less
than 1 TeV$^{-1}\sim 10^{-16}$~mm. On the other hand, the
four-dimensional character of gravity has been tested only down to the
centimeter scale. This sets un upper bound on the size $L$ of the extra
dimensions\footnote{This is in the simplest models, where the extra
dimensions are taken to be flat. In models such as the one of Randall
and Sundrum \cite{rRS}, the bound is on the curvature of the extra
dimensions.}. Since the effective four-dimensional Newton's constant
$G_4$ is related to its $d$-dimensional (fundamental) counterpart $G_d$
by $G_4 = G_d/L^{d-4}$, if the fundamental scale of gravity in the bulk
is of order a TeV, and we take $d=6$, then $G_4$ has the observed value
provided $L\sim 1$~mm, consistent with the bound mentioned above, and
falsifiable in the near future. This value of $L$ can be lowered, while
keeping a fundamental TeV energy scale, by taking higher values for $d$,
which leads to smaller extra dimensions. In what follows, our results
will hold for any number of large extra dimensions.

For an observer that lives on the brane ({\it e.g.,}\ ourselves), the main
effect of these large extra dimensions is to introduce a number of light
(and heavy) fields, which come from the decomposition of the bulk metric
into the four-dimensional graviton and an infinite tower of Kaluza-Klein
modes. The latter act like four-dimensional spin-two fields with masses
that, for $d=6$, start at as low as $1/L\sim 10^{-4}$~eV. So, at a given
energy scale $E <$ TeV, the light fields on the brane consist of the
Standard Model fields, a four dimensional graviton (the zero mode in the
Kaluza-Klein decomposition of the $d$-dimensional graviton), and a large
number, of order $(EL)^{d-4}$, of light Kaluza-Klein modes. The latter,
however, couple very weakly, with four-dimensional gravitational
strength, to the matter fields that are confined on the brane.

The existence of a low fundamental Planck scale implies that the
strength of gravity becomes comparable to other interactions at around
the TeV scale. One of the most striking consequences of this is the
possibility of forming semiclassical black holes at rather low energies,
say of order 100~TeV. Suppose one collapses matter (or collides
particles) on the brane to form a black hole of size $\ell_{\rm fun} \ll
r_0 \ll L$ (where $\ell_{\rm fun}=G_d^{1/(d-2)}$ is the fundamental, {\it
i.e.,}\ $d$-dimensional, Planck length). This black hole has a
temperature $T\sim 1/r_0$ which is much larger than the mass of the
light Kaluza-Klein modes. Since gravity couples to everything, and there
are so many Kaluza-Klein modes with mass less than the Hawking
temperature, it has been claimed \cite{rADM,rBF} that the Hawking
radiation will be dominated by these Kaluza-Klein modes, with only a
tiny fraction of the energy going into standard model particles. In
other words, most of the energy would be radiated off of the brane into
the bulk. Since the Kaluza-Klein modes couple so weakly to matter, they
would escape our detectors and therefore, if this argument were
correct, the Hawking radiation from these small black holes would be
essentially unobservable. 

However, we have proven, in work with Gary Horowitz and Rob Myers
\cite{rEHM3}, that this argument is incorrect, and most of the
Hawking radiation goes into the Standard Model fields on the brane.
While the detection of this Hawking radiation would likely not be the
first experimental signature of large extra dimensions, such
measurements would provide a dramatic new window on black hole
microphysics.

\section{Black Holes on Branes}

We start by reviewing the description of black holes in a brane-world
with extra dimensions of size $L$. We are far from having any exact,
analytic description that accounts for all the effects involved. In
fact, such a detailed description would be strongly model-dependent, but
if we make several approximations we will be able to obtain generic
results in the regimes of most interest. First, we will take the brane
to have negligible thickness. This is indeed reasonable, since the
actual thickness of the brane is likely to be of order the fundamental
scale $\ell_{\rm fun}$, and a black hole will behave semi-classically
only if its size is $r_0 \gg \ell_{\rm fun}$. Therefore, we will work in
a sort of low energy effective approach where we do not probe the
details of the structure of the brane. The latter is simply represented
as a four-dimensional Lorentzian hypersurface in the full spacetime,
and, in principle, acts as a distributional source for gravity. 

Even in this approximation, it has not been possible to find a full
analytic description of a black hole on a brane in any realistic model,
although it has been possible to work out in detail a useful toy model
in low dimensions \cite{rEHM,rEHM2}. In general, the self-gravity
of the brane introduces severe complications, and makes the
analysis strongly dependent on the number of extra dimensions. When
describing large black holes (relative to the compactification scale;
see below) this difficulty can be easily overcome, but the
representation of small black holes proves still too hard in general.
Small black holes are in fact much more effective at probing the extra
dimensions, and will be the main focus here. If we take the horizon size
to be sufficiently smaller than the length scale of the curvature
induced by the brane, then it will be reasonable to neglect brane
self-gravity. Also, if the black hole size is much smaller than the
compactification length, then, close to the black hole, finite-size
effects will be negligible, while at distances much larger than $L$ we
will be able to integrate over the internal space. In this way our
results will be largely independent of the precise compactification
scheme. Nevertheless, for definiteness and simplicity, we will be mainly
considering the extra dimensions to be wrapped on a square
$(d-4)$-dimensional torus. 

Let us then consider a general dimension $d$ for the bulk spacetime, and
assume that we live on a (3+1)-dimensional brane. A black hole horizon
on a three-brane arises from the intersection of a higher dimensional
horizon with the brane worldvolume. This horizon can originate from two
different sorts of
higher dimensional objects: one is the ``black brane,'' (in fact, a
$(d-4)$-brane) obtained by
taking the product of the four dimensional Schwarzschild
solution and the $(d-4)$-dimensional internal space ${\cal M}_{d-4}$,
\begin{equation}\label{bbrane}
ds^2=ds^2_{{\rm Schwarzschild}_4}+ds^2_{{\cal M}_{d-4}}\ .
\end{equation}
An observer on the brane (which is localized at a certain point
in the internal space) perceives exactly the four-dimensional
Schwarzschild solution, without any corrections arising from the
existence of extra dimensions.
In other words, no Kaluza-Klein modes are excited in this solution, only
the massless zero mode of the bulk graviton that yields four dimensional
gravity on the brane.

On the other hand, one can also envisage a different configuration that
results in a horizon on the brane, namely, a $d$-dimensional black hole
intersected by the three-brane. This is a localized black
hole, as opposed to the black brane which is delocalized in the extra
dimensions. Now, we are taking the extra dimensions to be compactified
with characteristic length $L$. In the simplest case of toroidal
compactification this is equivalent to regarding space in the directions
transverse to the brane as being periodic with period $L$. While this
periodicity can be readily imposed on the black brane solution above, a
localized black hole will instead require for its description a
$(d-4)$-dimensional array of $d$-dimensional black holes. No such exact
periodic solutions are known in dimensions $d>4$. Nevertheless, if we
restrict to black holes of size $r_0$ much less than $L$, then the
geometry {\it near the black hole} will be very well approximated by a
$d$-dimensional Schwarzschild solution,
\begin{equation}\label{metric}
ds^2=-f(r)\,dt^2+f^{-1}(r)\,dr^2
+r^2d\Omega_{d-2}^2
\end{equation}
with $f(r)=1-(r_0/ r)^{d-3}$.
The event horizon is 
at $r=r_0$, and 
has area $A_d=r_0^{d-2}\Omega_{d-2}$
where $\Omega_n$ denotes the volume of a unit $n$-sphere. 
Since we are neglecting its self-gravity, the brane is to be identified
as a surface of vanishing extrinsic curvature,
which, by symmetry, cuts through the equator of the black hole.
Then the induced metric on the brane will be
\begin{equation}\label{inmetric}
ds^2=-f(r)\,dt^2+f^{-1}(r)\,dr^2
+r^2d\Omega_2^2\ .
\end{equation}
On the brane then, the event horizon
is again at $r=r_0$, and its area 
is $A_4=4\pi
r_0^2$. This induced metric on the brane is certainly not the
four-dimensional Schwarzschild geometry. Indeed, one can think of it as
a black hole with matter fields ({\it i.e.,}\ Kaluza-Klein modes) around
it. However, the calculation of Hawking evaporation relies mainly on
properties of the horizon, such as its surface gravity ({\it i.e.,}\
temperature).
Since the
Hawking temperature is
constant over the horizon, it is the same for both the black hole in the
bulk and on the brane, and is given by $T=(d-3)/( 4\pi r_0)$.

In the metric (\ref{inmetric}) the $g_{tt}$ component
has no $1/r$ term and hence seems to give zero mass in four dimensions.
However, this metric only describes the geometry near the black hole. In
order to consider the metric at distances large from the black hole we
have to take into account the effects of compactification, {\it i.e.,}\ of the
full array of black holes, each of mass $M$ and separated by a distance
$L$. From a large distance\footnote{Here, we ignore the gravitational
interaction energy of the black holes in the array, which is justified
for $r_0\ll L$.},
the periodic array looks like a ``surface density" $\rho =M/L^{d-4}$,
where $M$ is the mass of the $d$-dimensional black hole.
Thus, asymptotically the metric will be of the form (\ref{inmetric}),
but now with
$f(r) = 1- (2G_d
\rho/r)$. However, since $G_d = G_4 L^{d-4}$, this is equivalent to
$f(r) = 1- (2G_4 M/r)$. So, for $r\gg L$
the geometry will be approximated by (\ref{inmetric}) with 
$f(r)\simeq 1-(2G_4 M/r)$, and the mass measured on the brane
is the same as the mass in the bulk. 

So we have two objects, namely, the extended brane and the localized
black hole, that can describe a black hole of size $r_0$ on the
brane\footnote{However, in the Randall-Sundrum model with a non-compact
dimension the extended solution---a black string---is unphysical
\cite{rCHH}, and both large and small black holes are localized on the
brane. The above description of small black holes is still essentially
valid, but large black holes are rather like pancakes on the brane
\cite{rEHM,rGKR}.}. Which of the two is the preferred configuration for
a collapsed object? The answer can be easily determined on the basis of
entropy arguments, which are furthermore supported by a study of the
classical stability of the black brane \cite{rGL}. It turns out that,
for $r_0$ greater than the compactification size $L$, the black brane
has larger entropy and thus dominates. On the other hand, for $r_0$
roughly smaller than $L$, an instability of the black brane sets in so
that the localized, higher dimensional black hole becomes the stable
(and more entropic) configuration. The transition between these two
regimes is very poorly understood, and any progress in the construction
of periodic arrays of black holes in dimensions larger than four would
be of great help.

Large black holes, therefore, give us virtually no clues as to the
presence of extra dimensions. It is by looking at small black holes that
we can expect to probe the physics of large extra dimensions. We now
turn to study their evaporation through Hawking radiation.

\section{Radiation On and Off the Brane}

Since we are interested in black holes of radius much less than the
size of the internal space, we can treat, as far as the radiation
process is concerned, the extra dimensions as non-compact. 
Then, for a single massless bulk field, the rate at which
energy is radiated is of order
\begin{equation}
\label{hawsix} {dE\over dt} \sim A_d\, T^d \sim {r_0^{d-2} \over r_0^d} 
\sim {1\over r_0^2} 
\end{equation}
where $A_d$ denotes the area of the higher dimensional black hole. For a
single massless four-dimensional field on the brane, the rate of energy
loss is of order
\begin{equation}\label{hawfour}
 {dE\over dt} \sim A_4 \,T^4 \sim {r_0^2 \over r_0^4} 
\sim {1\over r_0^2} 
\end{equation}
and hence is the same. That is, with a single relevant scale $r_0$
determining the Hawking radiation, bulk and brane fields must both have
$dE/dt\sim r_0^{-2}$. Hence the Hawking evaporation must emit comparable
amounts of energy into the bulk and brane. However, with the typical
assumption that there are many more fields on the brane than in the
bulk, one would conclude that most of the energy goes into the
observable four-dimensional fields. 

Notice that it would be incorrect to think of brane fields as bulk fields
confined to a limited phase space. The brane fields are intrinsically
four-dimensional, and their emission is governed by the four-dimensional
relation (\ref{hawfour}), and {\it not} the $d$-dimensional formula
(\ref{hawsix}) with a restricted area.

Another important point worth stressing is that even if there are a
large number (of order $(L/r_0)^{d-4}$) of light Kaluza-Klein modes with
masses below the
scale of the Hawking temperature, they do {\it not} dominate the evaporation.
The pitfall here is to think
of the individual Kaluza-Klein modes of the bulk graviton as massive
spin two fields on the brane with standard (minimal) gravitational
couplings. Rather, since the Kaluza-Klein modes are excitations in the
full transverse space, their overlap with the small ($d$-dimensional)
black holes is suppressed by the geometric factor $(r_0/L)^{d-4}$ relative
to the brane fields. Hence this geometric suppression precisely
compensates for the enormous number of modes, and the total contribution
of all Kaluza-Klein modes is only the same order as that from a single
brane field. 

In order to see in more detail how this geometric suppression factor
appears, it is instructive to look into the calculation of the emission
rate of a massless bulk field from the four dimensional perspective,
{\it i.e.,}\ by decomposing it into Kaluza-Klein modes. 
Thus, let us
separate the modes of the bulk field according to the momentum ${\bf k}$
which they carry into the $d-4$ transverse dimensions. On the brane,
this
Kaluza-Klein momentum is identified with the {\it four-dimensional} mass
of these modes, which we denote $m=|{\bf k}|$. If we then sum over all
other quantum numbers, we will find the emission rate corresponding to a
Kaluza-Klein mode with momentum ${\bf k}$. In this way, we
get, for the emission rate per unit frequency interval, of modes with
momenta in the interval (${\bf k},
{\bf k}+ d{\bf k}$),
\begin{equation}\label{ratekk}
{dE\over d\omega dt}(\omega,{\bf k})\simeq
(\omega^2-m^2) {\omega A_d
\over e^{\beta\omega}-1}\,d^{d-4}\!k\ .
\end{equation}
Here, $A_d$ is the area of the black hole in the $d$-dimensional
bulk.
We are neglecting purely numerical factors since we
have found that they do not play any significant role. 

Consider a light Kaluza-Klein mode, with a mass
much smaller than the black hole temperature, 
$m \ll 1/r_0$. 
We set $d^{d-4}\!k\sim (1/L)^{d-4}$ for an individual mode, and 
$A_d\sim r_0^{d-4} A_b$, with $A_b$ the sectional area on the brane.
Then, 
\begin{equation}\label{lightkk}
{dE\over d\omega dt}(\omega,m)\simeq \left( r_0\over L\right)^{d-4}
(\omega^2-m^2) {\omega A_b\over
e^{\beta\omega}-1}\,.
\end{equation}
which is identical to the emission rate of a massive field in four
dimensions, except for a suppression factor of $(r_0/L)^{d-4}$. (Note
that this formula applies equally well for $m=0$.) So we see that the
Hawking radiation into each Kaluza-Klein mode (among these, the massless
graviton) is much smaller, by a factor of
$(r_0/L)^{d-4}$, than the radiation into any other minimally
coupled field that propagates only in four dimensions. Still the total
radiation into a bulk field is comparable to that into a field on the
brane, because there are of order $(L/r_0)^{d-4}$ light modes with
$m<T\sim1/r_0$. So if we integrate the emission rate over all
Kaluza-Klein modes, and then over frequency, we recover
eq.~(\ref{hawsix}), as it must be for consistency.

Since the number of relevant fields on the brane may be only a factor of
ten or so larger than the number of bulk fields, one might worry that
the claim that the Hawking radiation is dominated by brane fields could
still be thwarted by large numerical factors coming from the higher
dimensional calculation. To check this, we have considered two improvements
over the rough estimate of the radiation rates given in (\ref{hawsix}) and
(\ref{hawfour}). The first is to include the dimension dependent
Stefan-Boltzman constant that appears in the black body
radiation formula. A second improvement concerns the effective area of the
radiating black body, which is slightly larger than the horizon area. We
have found there
are no unexpected large factors to ruin the
naive estimate that a Hawking evaporation emits as much energy
into a typical brane field as into a typical bulk field. A definitive
comparison of the bulk and brane radiation rates would require a detailed
analysis, with a
specific brane-world model to determine the exact black hole geometry
and the precise multiplicity of bulk and brane fields. 

So far we have considered small black holes with $r_0 < L$. Larger black
holes (which are described by (\ref{bbrane})) also radiate mainly on the
brane. The essential feature now is that the Hawking temperature is
lower than the mass of all Kaluza-Klein modes, so their contribution to
the Hawking radiation is clearly suppressed. Also, the massless mode of
the bulk field radiates in this regime identically to a brane field. So
for large black holes, a bulk field still carries essentially the same
energy as a field on the brane, but the latter again dominate the
Hawking radiation since they are more numerous.

We have also shown in \cite{rEHM} that black holes in the
Randall-Sundrum scenario \cite{rRS} with an infinite extra dimension still
radiate mainly on the brane. 

\section{The Evaporation Process}

We are thus led to the following picture for the evaporation of a black
hole that initially has a radius $r_0>L$. At the beginning, as a result
of Hawking evaporation, it will decrease its size, at a rate determined
by four dimensional physics. When $r_0\sim L$, the solution
(\ref{bbrane}) becomes unstable \cite{rGL}, and is believed to break up
into $d$-dimensional black holes which coalesce and form a single higher
dimensional black hole. It can be shown \cite{rEHM3} that this final
black hole will lie in the brane, and not in the bulk, since it feels a
restoring force due to the brane tension. From this point on, the
evaporation of the small black hole starts to differ from the way
predicted by the four dimensional law. The radiation rate, as we have
seen, varies smoothly across the transition between regimes, and the
evaporation slows down. This is a consequence of the fact that, in this
regime, the black brane has given way to another configuration, the
localized black hole, that, for a given mass, has higher entropy.
Although the area is larger, the temperature is lower, and the
evaporation rate is slower. As a consequence, the lifetime of a small
black hole is longer (possibly enormously longer) by a factor
$(L/r_0)^{2(d-4)}$ than would have been expected from four-dimensional
Einstein gravity. Finally, notice that the black hole can be described
semiclassically down to a mass scale (say, of order 100~TeV) much
smaller than the four dimensional Planck mass of $10^{19}$~GeV
\cite{rBF}.

So we conclude that the brane-world scenario has the potential to make
interesting observable predictions about small black holes appearing
either in collider experiments or in the early universe. Much of their
detailed phenomenology is still to be investigated.

\acknowledgments

I would like to thank the organizers, and in particular Bernard Julia,
for the invitation to present this work at this TMR conference. I would
also like to thank Gary Horowitz and Rob Myers, with whom the results
reported here were obtained. Work supported by UPV grant
063.310-EB187/98 and CICYT AEN99-0315. Attendance to the conference was
partially supported by TMR program under EC contract number FMRX-CT96-
0012.

\end{document}